\documentclass{emulateapj}
\shorttitle{Early Afterglow of GRB 030418}
\shortauthors{Rykoff et al.}

\begin{document}

\slugcomment{Accepted to Astrophysical Journal Letters, 16 October 2003}

\title{The Early Optical Afterglow of GRB~030418 and Progenitor Mass Loss}

\author{Rykoff,~E.~S.\altaffilmark{1}, Smith,~D.~A.\altaffilmark{1},
Price,~P.~A.\altaffilmark{2}, Akerlof,~C.~W.\altaffilmark{1},
Ashley,~M.~C.~B.\altaffilmark{4}, Bizyaev,~D.\altaffilmark{5,6},
Garradd,~G.~J.\altaffilmark{3}, McKay,~T.~A.\altaffilmark{1},
McNaught,~R.~H.\altaffilmark{3}, Phillips,~A.\altaffilmark{4},
Quimby,~R.\altaffilmark{7}, Schaefer,~B.\altaffilmark{8},
Schmidt,~B.\altaffilmark{3}, Vestrand,~W.~T.\altaffilmark{9},
Wheeler,~J.~C.\altaffilmark{7}, Wren,~J.\altaffilmark{9} }

\altaffiltext{1}{University of Michigan, 2477 Randall Laboratory,
	500 E. University Ave., Ann Arbor, MI, 48104, erykoff@umich.edu,
	donaldas@umich.edu, akerlof@umich.edu, tamckay@umich.edu}
\altaffiltext{2}{Institute for Astronomy, University of Hawaii, Honolulu, HI
	96822, price@ifa.hawaii.edu}
\altaffiltext{3}{Research School of Astronomy and Astrophysics, Mount Stromlo
	Observatory, via Cotter Road, Weston, ACT 2611, Australia,
	gloomberah@ozemail.com.au, rmn@murky.anu.edu.au, brian@mso.anu.edu.au}
\altaffiltext{4}{School of Physics, Department of Astrophysics and Optics,
	University of New South Wales, Sydney, NSW 2052, Australia,
	mcba@phys.unsw.edu.au, a.phillips@unsw.edu.au}
\altaffiltext{5}{Department of Physics, University of Texas at El Paso, El
	Paso, TX 79968, dmbiz@baade.physics.utep.edu}
\altaffiltext{6}{Sternberg Astronomical Institute, Moscow, Russia}
\altaffiltext{7}{Department of Astronomy, University of Texas, Austin, TX
	78712, quimby@astro.as.utexas.edu, wheel@astro.as.utexas.edu}
\altaffiltext{8}{Department of Physics and Astronomy, Louisiana State
	University, Baton Rouge, LA 70803, brad@baton.phys.lsu.edu}
\altaffiltext{9}{Los Alamos National Laboratory, NIS-2 MS D436, Los Alamos, NM
	87545, vestrand@lanl.gov, jwren@nis.lanl.gov}
\begin{abstract}

The ROTSE-IIIa telescope and the SSO-40 inch telescope, both located at Siding
Spring Observatory, imaged the early time afterglow of GRB~030418.  In this
report we present observations of the early afterglow, first detected by the
ROTSE-IIIa telescope 211~s after the start of the burst, and only 76~s after
the end of the gamma-ray activity. We detect optical emission that rises for
$\sim600\,\mathrm{s}$, slowly varies around $R=17.3\,\mathrm{mag}$ for
$\sim1400\,\mathrm{s}$, and then fades as a power law of index $\alpha=-1.36$.
Additionally, the ROTSE-IIIb telescope, located at McDonald Observatory, imaged
the early time afterglow of GRB~030723.  The behavior of this light curve was
qualitatively similar to that of GRB~030418, but two magnitudes dimmer.  These
two afterglows are dissimilar to other afterglows such as GRB~990123 and
GRB~021211.  We investigate whether the early afterglow can be attributed to a
synchrotron break in a cooling synchrotron spectrum as it passes through the
optical band, but find this model is unable to accurately describe the early
light curve. We present a simple model for gamma-ray burst emission emerging
from a wind medium surrounding a massive progenitor star.  This model provides
an effective description of the data, and suggests that the rise of the
afterglow can be ascribed to extinction in the local circumburst environment.
In this interpretation, these events provide further evidence for the
connection between gamma-ray bursts and the collapse of massive stars.
\end{abstract}

\section{Introduction}

Around half of all well-localized gamma-ray bursts (GRBs) have resulted in the
detection of optical counterparts.  This low success rate is partly due to the
difficulty in obtaining prompt coordinates and so it has been argued that many
GRB afterglows fade too rapidly for discovery by late time follow-up
observations. It is also possible that extinction from dense circumburst
environments may cut optical emission below detectable levels~\citep{khghc03}.
GRB~990123 remains unique as the only burst from which prompt optical emission
was detected during gamma-ray emission~\citep{abbbb99}, despite much effort
from small rapidly responding telescopes such as ROTSE-I and
LOTIS~\citep{abbbb00,kabbb01,ppwab99}. Larger fast-slewing telescopes such as
ROTSE-IIIa have since come online in an effort to achieve deeper imaging at
early times.  To date, only two other afterglows have been detected within 10
minutes of the burst---GRB~021004~\citep{fyktk03} and
GRB~021211~\citep{fpsbk03,lfcj03}---and both of these were detected only after
the afterglow began to decay.

In this paper, we report on early-time optical observations of GRB~030418 with
the ROTSE-IIIa (Robotic Optical Transient Search Experiment) telescope and the
SSO 40-inch telescope, both located at Siding Spring Observatory, Australia.
We also report on early-time optical observations of GRB~030723 with the
ROTSE-IIIb telescope at McDonald Observatory, Texas.  Despite rapid responses
to each of these bursts (211~s and 47~s respectively), we have no evidence of
prompt optical counterparts.  We present here a physical model that ascribes
the afterglow rise to extinction in the local circumburst environment.

The ROTSE-III array is a worldwide network of 0.45~m robotic, automated
telescopes, built for fast ($\sim 6$ s) responses to GRB triggers from satellites
such as HETE-2.  They have a wide ($1\fdg85 \times 1\fdg85$) field of
view imaged onto a Marconi $2048\times2048$ back-illuminated thinned CCD, and
operate without filters.  The ROTSE-III systems are described in detail in
\citet{akmrs03}.

The SSO 40-inch telescope has an f/8 direct imager at a Cassegrain focus.  The
field of view has a $20.8\arcmin$ diameter on a Tek $2048\times2048$ CCD with
24 micron pixels.  The telescope can be operated unfiltered or with a range of
filters.  For these observations the CCD was used with $2\times2$ binning
giving pixels of $1\farcs2$.

\section{Observations and Analysis}
On 2003 April 18, HETE-2 detected GRB~030418 (HETE-2 trigger 2686) at
9:59:18.85 UT.  The first determination of its position was distributed as a
Gamma-ray Burst Coordinates Network (GCN) notice at 10:02:54 UT, with a
$28\arcmin$ radius error box, 205~s after the start of the burst (HETE-2 2686;2).
Ground analysis improved the error radius to $18\arcmin$, and a second GCN
notice was distributed at 11:43:01 UT (HETE-2 2686;4).  The burst was
determined to have lasted 135~s, with a fluence of $1.2\times10^{-6}\,
\mathrm{erg}\,\mathrm{cm}^{-2}$ (2--25 keV) and
$2.5\times10^{-6}\,\mathrm{erg}\,\mathrm{cm}^{-2}$ (30--400 keV), classifying
it as a long, X-ray rich GRB~\citep{sgdms03}.  The burst occurred while the
Moon was bright (96\% illumination) and very few telescopes reported follow-up
observations.

ROTSE-IIIa responded automatically to the first GCN notice in under 6~s
with the first exposure starting at 10:03:00 UT, 211~s after the burst
and only 76~s after the cessation of gamma-ray activity.  The automated
scheduling software began a program of ten 5-s exposures followed by 90
20-s exposures.  Longer exposures were not taken because the bright Moon
would have saturated the images.  The second GCN notice triggered ROTSE-IIIa to
repeat the same sequence of observations.  Analysis of the
individual frames in near real-time did not reveal any new source brighter than
the unfiltered limiting magnitude of $\sim 16$.

After the receipt of the second GCN notice, \citet{pmgvs03} initiated a burst
response on the SS0 40-inch telescope, beginning 7139~s after the burst,
with an unfiltered 100-s image, followed by a sequence of 10 300-s images taken
with a Johnson $R$-band filter. \citet{pmgvs03} identified a new object at
$\alpha=10^h54^m33\fs674$, $\delta=-7\arcdeg01\arcmin40\farcs75$ (J2000.0), at
$\sim18.8$ magnitude that was not on the Digitized Sky Survey red
plates~\citep{pmgvs03}.

\begin{figure}
\scalebox{1.2}{\plotone{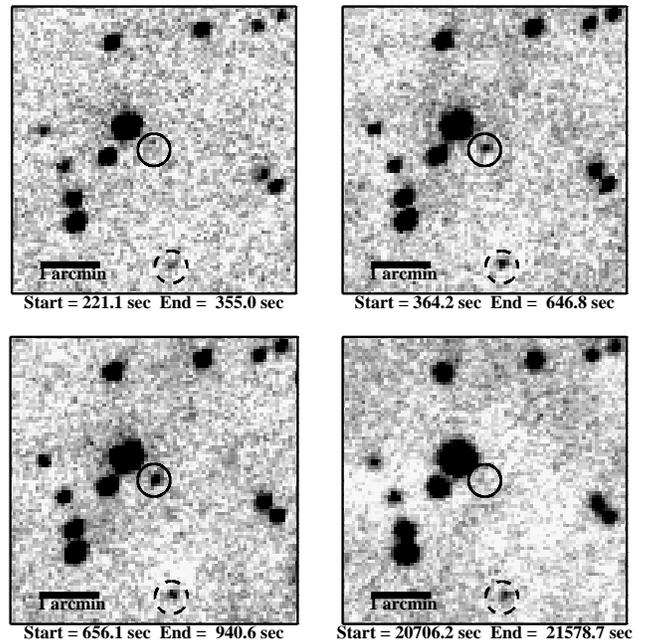}}
\caption{\label{fig:mosaic}Mosaic of co-added images taken by ROTSE-IIIa
	starting 211~s after the burst clearly show the rise and fall of the
	afterglow of GRB~030418.  In the first image the $18^{\mathrm{th}}$ magnitude
	comparison star (dotted circle) is clearly visible, while the afterglow is not.
	In subsequent images the afterglow (solid circle) gets brighter than the
	comparison star and fades away.  The four images have different total
  	exposure times, and the comparison star varies by $<0.1$ mag.}
\end{figure}

Co-adding sets of 10 ROTSE-IIIa images revealed the optical counterpart
reported by \citet{pmgvs03}.  As can be seen in Figure~\ref{fig:mosaic}, the
afterglow is barely detected in the first ROTSE-IIIa image, while a nearby
$18^{\mathrm{th}}$ magnitude star at $\alpha=10^h54^m32\fs6$,
$\delta=-07\arcdeg03\arcmin38\farcs0$ (J2000.0) is clearly visible.  The
afterglow then increases in brightness, exceeding the comparison star, before
fading below our detection threshold.  Using an identical analysis as for the
afterglow itself (to be described below), we measured the fluctuations of the
comparison star to be $<0.1$ magnitudes.  The comparison star is detected with
a S/N of $7.7$ in the first images, and a S/N of $>10$ in subsequent images.

\begin{figure*}
\rotatebox{270}{\scalebox{0.8}{\plotone{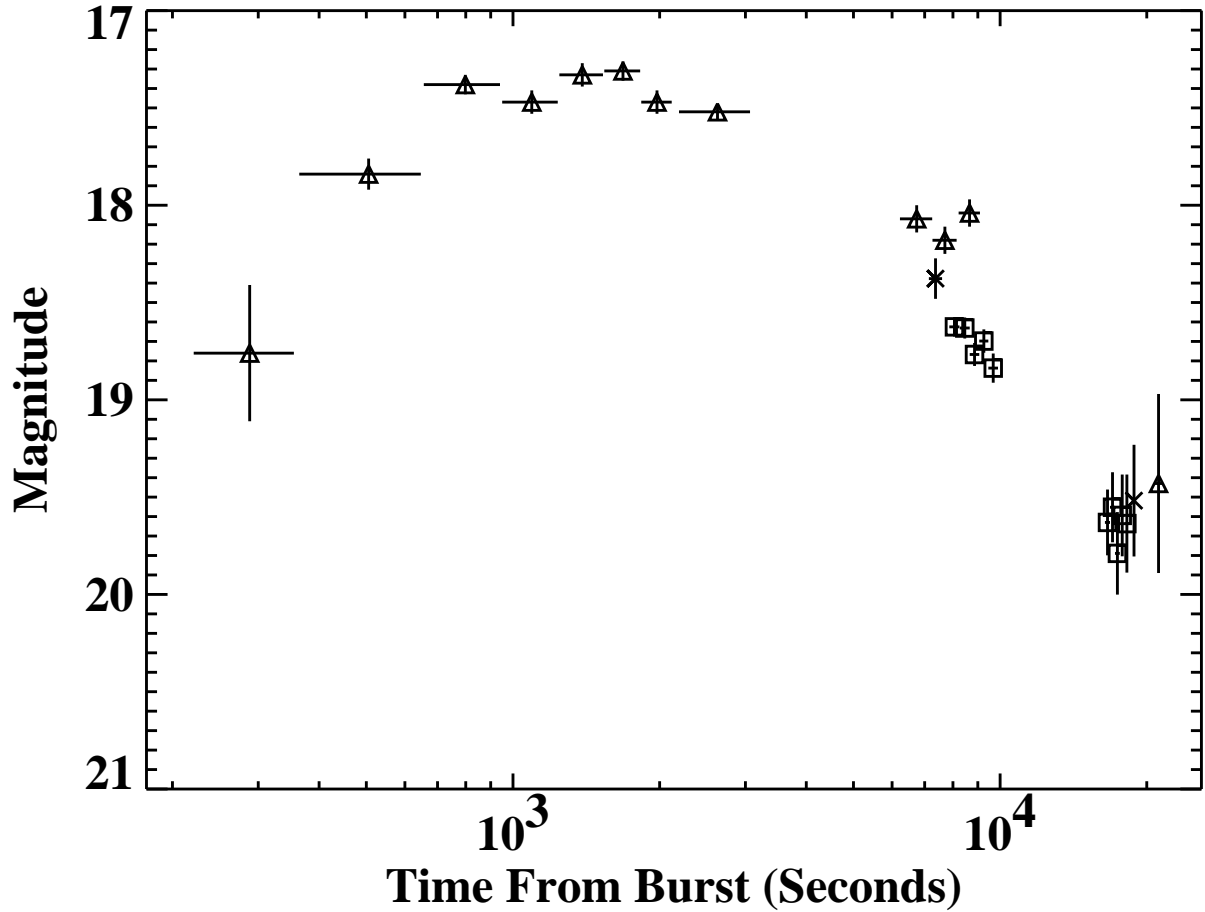}}}
\figcaption{\label{fig:lightcurve}Early-time light curve for GRB~030418.
  The optical emission rises during the first $600\,\mathrm{s}$, slowly varies
  for $1400\,\mathrm{s}$, and then fades as a power law.
  The triangles are unfiltered ROTSE-IIIa data, the crosses are unfiltered SSO
  40-inch data, and the squares are $R$-band SSO 40-inch data.}
\end{figure*}

The ROTSE-IIIa data and the first SSO 40-inch observation were taken without
filters.  Further SSO 40-inch observations were taken with an $R$-band filter.
Therefore, it is important to bring both sets of measurements to the same
standard photometric system.  We compared each image to the standard $R$-band
photometric calibration from the USNO-1m telescope~\citep{h03}.  Unfortunately,
we do not have any color information for the afterglow at the early times.
Table~\ref{tab:one} shows the results of our photometry for the afterglow, with
the light curve plotted in Figure~\ref{fig:lightcurve}.

\begin{deluxetable}{llccl}
\tablewidth{0pt}

\tablecaption{Optical Photometry for GRB~030418\label{tab:one}}
\tabletypesize{\scriptsize}
\tablehead{
\colhead{Telescope} & \colhead{Filter} & \colhead{Start (s)} & \colhead{End (s)} &
\colhead{Magnitude}
}
\startdata
ROTSE-IIIa & None & 221.1 & 355.0 & $18.76\pm0.35$ \\
ROTSE-IIIa & None & 364.2 & 646.8 & $17.84\pm0.08$\\
ROTSE-IIIa & None & 656.1 & 940.6 & $17.38\pm0.05$\\
ROTSE-IIIa & None & 950.2 & 1235.8 & $17.47\pm0.06$\\
ROTSE-IIIa & None & 1245.2 & 1530.3 & $17.33\pm0.06$\\
ROTSE-IIIa & None & 1539.8 & 1824.1 & $17.31\pm0.05$\\
ROTSE-IIIa & None & 1833.4 & 2117.3 & $17.47\pm0.06$\\
ROTSE-IIIa & None & 2192.1 & 3066.0 & $17.52\pm0.04$\\
ROTSE-IIIa & None & 6228.5 & 7253.4 & $18.07\pm0.07$\\
ROTSE-IIIa & None & 7262.9 & 8141.1 & $18.18\pm0.07$\\
ROTSE-IIIa & None & 8217.7 & 9091.3 & $18.04\pm0.07$\\
ROTSE-IIIa & None & 20706.2 & 21578.7 & $19.43\pm0.46$\\
SSO 40-inch & None & 7142 & 7242 & $18.38\pm0.10$\\
SSO 40-inch & $R$ & 7867 & 8167 & $18.63\pm0.05$\\
SSO 40-inch & $R$ & 8264 & 8564 & $18.63\pm0.05$\\
SSO 40-inch & $R$ & 8664 & 8964 & $18.77\pm0.06$\\
SSO 40-inch & $R$ & 9062 & 9362 & $18.70\pm0.06$\\
SSO 40-inch & $R$ & 9460 & 9760 & $18.84\pm0.08$\\
SSO 40-inch & $R$ & 16415 & 16715 & $19.63\pm0.17$\\
SSO 40-inch & $R$ & 16813 & 17113 & $19.55\pm0.18$\\
SSO 40-inch & $R$ & 17212 & 17512 & $19.79\pm0.21$\\
SSO 40-inch & $R$ & 17611 & 17911 & $19.59\pm0.21$\\
SSO 40-inch & $R$ & 18009 & 18309 & $19.64\pm0.25$\\
SSO 40-inch & $R$ & 81401 & 85346 & $>21.5$\\
\enddata
\end{deluxetable}

The ROTSE-IIIa images were bias-subtracted and flat-fielded in the standard
way.  The flat-field image was generated from 30 twilight images.  We used
SExtractor~\citep{ba96} to perform the initial object detection, and to
determine the centroid positions of the stars.  We used a robust PSF fitting
code to measure the photometry.  We used six bright, well-measured stars from
the \citet{h03} list to fit our PSF, as well as to determine an $R$-band
magnitude zero-point.  The intrinsic variations in object color generate an RMS
dispersion of $0.23$ between our unfiltered ROTSE magnitudes and the
\citet{h03} $R$-band magnitudes.

The SSO 40-inch images were bias-subtracted and flat-fielded with a twilight
flat in the standard way.  We used weighted-aperture photometry and set the
magnitude zero point for each frame using between 30 and 90 \citet{h03} stars.
The systematic error estimated from comparison to the \citet{h03} stars is
about 0.25 magnitudes.

\begin{figure}
\rotatebox{270}{\scalebox{0.8}{\plotone{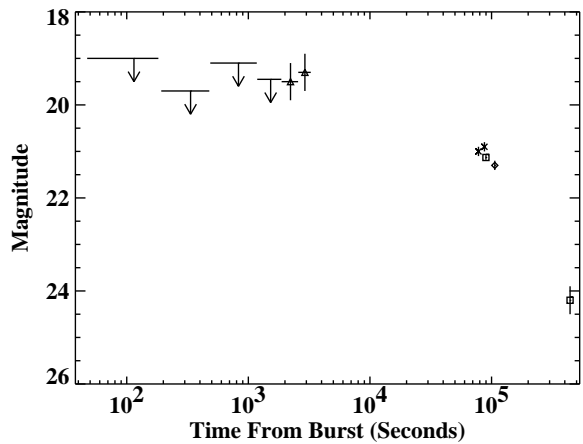}}}
\figcaption{\label{fig:0723lc}Light curve for GRB~030723.  The early time upper
  limits from ROTSE-IIIb imply behavior similar to GRB~030418.  The triangles and
  upper limits are ROTSE-IIIb data, the crosses are Palomar data~\citep{fkckn03},
  the squares are Magellan data~\citep{dbvvr03,dbvvr03b}, and the diamond is
  Cerro Tololo data~\citep{b03}.}
\end{figure}

On 2003 July 23, at 06:28:17.45 UT, HETE-2 detected another X-ray bright GRB
(HETE trigger 2777), 23~s in duration~\citep{pbcdd03}.  The ROTSE-IIIb
instrument responded automatically and began taking images within 5~s of the
GCN notice distribution.  The first ROTSE exposure began 47~s after the burst
trigger time.  The system took ten 5-s images, ten 20-s images, and 40 60-s
images of the burst field.  Much like GRB~030418, the burst counterpart was not
found in these early images to limiting magnitudes of 17--18 mag.  Later
observations by larger telescopes revealed a faint, fading source at
$\alpha=21^h49^m24\fs40$, $\delta=-27\arcdeg42\arcmin47\farcs4$ (J2000.0) at
$20^{\mathrm{th}}$ magnitude~\citep{fkckn03}.  We co-added sets of ten images
and applied the same PSF-fitting technique as described above.  The object was
not detected in our four earliest co-added images, but the last two images
yield marginal detections.  We derive $19.5\pm0.4$ for the fifth image
(S/N=2.7) and $19.3\pm0.4$ for the sixth image (S/N=3.1).  The resulting light
curve is shown in Figure~\ref{fig:0723lc}.

\section{Results}

Figure~\ref{fig:lightcurve} shows the early-time light curve of GRB~030418 from
the ROTSE-IIIa and SSO 40-inch observations.  After our first detection, the
afterglow brightness is rising.  After the rise, the afterglow slowly varies
around $17.3$ mag for about an hour, before fading following a power law.

Most GRB afterglows have been seen to decline with a sequence of one or more
power laws.  Two hours after the burst, this is also the case for GRB~030418.
Extrapolating this late time power law decay back to the early time provokes
the question of what happened to the missing optical flux.  There are two main
alternatives: intrinsically, there were fewer optical photons emitted by the
source or, extrinsically, the optical photons were absorbed after they were
created.  

\begin{figure}
\rotatebox{270}{\scalebox{0.87}{\plotone{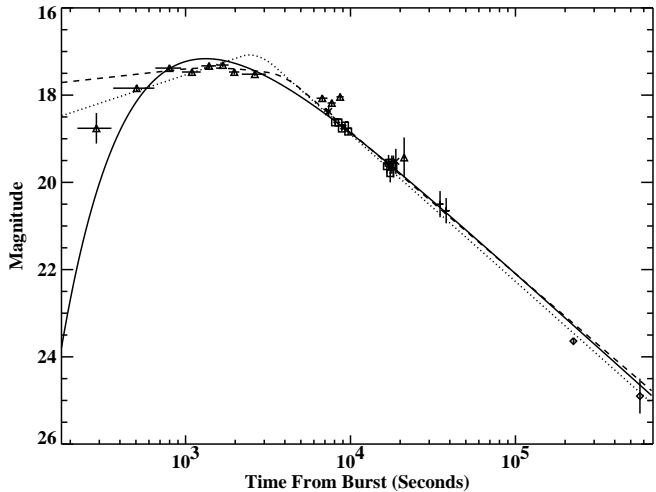}}}
\figcaption{\label{fig:lightcurvefit}Light curve for GRB~030418 with the
  best fit model superimposed. The solid line fit is of the form
  $F_\nu=F_0t^{-\alpha}e^{-\beta_t/t}$, where $\alpha=1.36\pm0.02$.  The dotted
  fit is a model with slow cooling as in \citet{spn98}, while the dashed
  line is a model with both slow and fast cooling. The triangles
  are ROTSE-IIIa observations, the crosses and squares are SSO 40-inch
  observations, the pluses are Loiano telescope data from \citet{fpbgp03}, and
  the diamonds are Magellan telescope data from \citet{dbrec03b,dbrec03}.}
\end{figure}

The first possibility is not easy to describe without making extensive
assumptions about the physics of the shock front or the density of the ambient
medium.  We have investigated the possibility that the light curve results from
the spectral break frequency coming through the optical band~\citep{spn98}.  We
modeled the flux spectrum as a two component power-law as in \citet{spn98} for
the slow cooling regime, with $F_\nu\propto\nu^{1/3}$ for $\nu<\nu_m$, and
$F_\nu\propto\nu^{-(p-1)/2}$ for $\nu>\nu_m$, where $\nu_m$ is the
characteristic synchrotron frequency, and $p$ is the spectral index of the
electrons, fit to the late time power-law decline.  The synchrotron frequency
decreases as $\nu_m\propto t^{-3/2}$, with a best fit value of $\nu_m =
9.7\times10^{14}\,\mathrm{Hz}$ at 1500~s after the burst.  After integrating
the flux in our optical passband, we found the predicted optical peak is much
too sharp and does not reproduce the smooth rollover we see in the light curve
of GRB~030418, as can be seen as the dotted line in
Figure~\ref{fig:lightcurvefit}.  This fast transition is not caused by the
sharpness of the frequency break, but rather the rapidity with which the break
moves through the optical band.

We next investigated the possibility that we are seeing the transition from
fast cooling to slow cooling, as in ~\citet{spn98}.  In addition to our model
described above, we added the cooling frequency, $\nu_c$, which decays as
$\nu_c\propto t^{-1/2}$.  By fitting the initial values of $\nu_c$ and $\nu_m$,
the modeled light curve is plotted as the dashed line in
Figure~\ref{fig:lightcurvefit}.  This model succeeds in describing the flat
peak of the light curve, but does not describe the rise at early times.  In
addition, the best fit frequency values at 1500~s for $\nu_c =
4.6\times10^{14}\,\mathrm{Hz}$ and $\nu_m = 1.8\times10^{15}\,\mathrm{Hz}$ are
not physically reasonable, according to the prescription in \citet{gs02}.

Recently, evidence has emerged linking GRBs to core-collapse supernovae of
massive stars, including the detection of a SN spectrum in the afterglow of
GRB~030329~\citep{smgmb03,hsmfw03}.  A key consequence of a massive star
progenitor is that the GRB occurs inside a massive stellar
wind~\citep{mrw98,cl99}.  To date, most of the literature has focused on the
generation of the shock front in a wind medium without consideration of other
physical effects.  Here, we investigate the possibility that the absorption of
the optical photons by this circumburst environment can explain the early
behavior of the GRB~030418 light curve.  Although a prompt optical/UV flash
might sublimate the dust~\citep{wd00}, if this were the entire story then we
would likely have seen the afterglow decay at the early time.

The stellar wind density profile is $\rho={\rho_0}r_0^2/r^2$,
assuming a constant mass loss rate from the massive star.  The optical depth
scale then becomes $\tau=\beta_r/r$, and the optical flux absorbed by the
circumburst medium is attenuated as $e^{-\beta_r/r}$.  If we assume that at the
early time the emitting shell is moving at a roughly constant velocity with a
bulk Lorentz factor of $\Gamma$, then the distance traveled is related to the
time in the Earth's frame of reference as $r\simeq\frac{1}{2}\Gamma^2ct$.
Therefore, the attenuation as a function of time goes as $e^{-\beta_t/t}$.  As
the standard model has the afterglow fading as a power law after the initial
energy injection into the circumburst medium, we assume that the power law
decline applies at all relevant times.  We can then fit the rise and fall of the
afterglow with an attenuated power law function,
\begin{equation}
 F_\nu = F_0 t^{-\alpha} e^{-\beta_t/t}\label{eqn:thefunction}.
\end{equation}
We derive the three free parameters of this fitting function, $F_0$, $\alpha$, and
$\beta_t$ empirically via a linear regression fit to the observed
afterglow light curve.  The best fit function is plotted in
Figure~\ref{fig:lightcurvefit}, with $\alpha=1.36\pm0.02$, and $\beta_t=1.81
(\pm0.05) \times10^{3}\,\mathrm{s}$, and $F_0$ is an arbitrary normalization
factor.  The formal errors quoted are from the linear regression fit.  The fit
has a $\chi^2/\mathrm{DOF}=8.7$.  This large value is mostly due to local
effects in the data at 8000~s.

From Figure~\ref{fig:lightcurvefit} it can be seen that this physical model is
able to represent the gross features of the light curve.  We will now show
that the observed value for the attenuation time-scale, $\beta_t$, is
consistent with reasonable assumptions about the circumburst medium.  To
estimate the mass loss implied by $\beta_t$, we assume that the primary
absorbing medium is dust grains similar in composition to the ISM.  The opacity
of the ISM can be approximately modeled as
$\kappa_\nu=(5\times10^{-13}\,\mathrm{cm}^2\mathrm{g}^{-1}\mathrm{s})
\nu$~\citep{dl84,as85}.  Although there is no redshift measurement for this
burst, we assume a typical redshift of $z\sim1$, which puts the peak response
of our CCD at $\lambda\sim 325\,\mathrm{nm}$ in the GRB local frame of
reference.

We estimate the wind velocity as
$v_{wind}\sim100\,\mathrm{km}\,\mathrm{s}^{-1}$, the escape velocity from a
$50M_\sun$ star with a radius $2000R_\sun$, a typical value for a red
supergiant.  Finally, we assume that the bulk Lorentz factor of the GRB ejecta
is 100.  The derived mass loss rate for the progenitor star is then:
\begin{equation}
  \frac{dM}{dt} = 1.2\times10^{-3} \frac{M_\sun}{\mathrm{yr}}
  \left(\frac{\Gamma}{100}\right)^2 \frac{\beta_t}{1800}
  \frac{450\,\mathrm{cm}^2\,\mathrm{g}^{-1}}{\kappa}
  \frac{v}{100\,\mathrm{km}\,\mathrm{s}^{-1}}\label{eqn:dmdt},
\end{equation}
where $\Gamma$ is the bulk Lorentz factor, $\kappa$ is the opacity and $v$ is
the velocity of the stellar wind.  With our assumptions, a mass loss rate of
$1.2\times10^{-3}\,M_\sun\mathrm{yr}^{-1}$ is high compared to that of late
stage, high mass stars, which is typically
$10^{-5}$--$10^{-4}\,M_\sun\mathrm{yr}^{-1}$~\citep{glm96, gml96}.  However,
such a mass loss rate may not be out of line for the extreme stellar masses
required by the prevailing collapsar/hypernova models of GRB progenitors.  This
estimate is also dependent on the accuracy of our assumed value for the bulk
Lorentz factor -- a factor of three lower would reduce the mass loss by a
factor of ten.  Additionally, our estimate is only reasonable with a mass loss
rate typical of a red supergiant, not a Wolf-Rayet star as discussed as a possible
progenitor for GRB~021004~\citep{sghpq03}.  

\section{Discussion}

GRB~030418 is one of the earliest afterglows yet imaged, with the initial
detection only 76~s after the cessation of the gamma-ray activity.
Unlike GRB~990123, this burst does not appear to have a prompt optical
counterpart that can be attributed to the reverse shock.  However, if our model
of local extinction is correct, we would not expect to see any prompt emission;
this model implies optical extinction of roughly 20 magnitudes at
$100\,\mathrm{s}$, near the time gamma-ray emission ceased.

A backward extrapolation of the late power law decline overestimates the
optical emission from GRB~030418.  Our model of local dust absorption in a
stellar wind medium is a useful way of characterizing the data.  Unlike
frequency break models, our model is able to describe the steep rise and slow
rollover of the light curve.  This early time behavior is far from universal,
as several bursts, including GRB~990123, and GRB~021211, had more emission than
predicted from the late power law decline.  However, we already have evidence
that the light curve behavior of GRB~030418 is not unique.  Our early-time
observations of GRB~030723 show a similarity to the light curve of GRB~020418.
Given the observational biases against detecting such dim fading objects, it is
not surprising that this class of GRB afterglows is just now being discovered
as a consequence of more accurate coordinate determinations in space and more
sensitive optical detectors on the ground.

One of the main consequences of our absorption model in a stellar wind medium
is that some afterglows will rise very steeply in the early time.  It is at
this very early time that the degeneracy between our model and the frequency
break models is broken.  Another consquence of a dusty local environment is
that the extinction in the optical bands will be much greater than in the near
infrared.  Prompt multi-color observations will therefore be invaluable to
firmly establish if this type of initial behavior is due to optical absorption
as described above.

\acknowledgments 

This work has been supported by NASA grants NAG5-5281 and
F006794, NSF grants AST-0119685 and 0105221, the Australian Research Council,
the University of New South Wales, and the University of Michigan.  Work
performed at LANL is supported by NASA SR\&T through Department of Energy (DOE)
contract W-7405-ENG-36 and through internal LDRD funding.  Special thanks to
the staff at Siding Spring and McDonald Observatories, to Scott Barthlemy for
maintaining the GCN and to Fred Adams for his illuminating discussion of dust
absorption.

\newcommand{\noopsort}[1]{} \newcommand{\printfirst}[2]{#1}
  \newcommand{\singleletter}[1]{#1} \newcommand{\switchargs}[2]{#2#1}

\end{document}